\documentclass[letterpaper,twocolumn,10pt]{article}
\usepackage{usenix,epsfig,endnotes}
\usepackage[pdftex]{color}
\usepackage{gnuplot-lua-tikz}
\usepackage{url}

\begin{document}

\date{5 September 2013}

\title{Optimize Unsynchronized Garbage Collection in an SSD Array}

\author{
{\rm Da Zheng, Randal Burns}\\
Department of Computer Science\\
Johns Hopkins University\\
\and
{\rm Alexander S. Szalay}\\
Department of Physics and Astronomy\\
Johns Hopkins University\\
} 

\maketitle

\thispagestyle{empty}

\subsection*{Abstract}
Solid state disks (SSDs) have advanced to outperform traditional hard
drives significantly in both random reads and writes. However, heavy
random writes trigger frequent garbage collection and decrease the performance
of SSDs. In an SSD array, garbage collection of individual SSDs is not
synchronized, leading to underutilization of some of the SSDs.

We propose a software solution to tackle the unsynchronized garbage collection
in an SSD array installed in a host bus adaptor (HBA), where individual
SSDs are exposed to an operating system. We maintain a long I/O queue for
each SSD and flush dirty pages intelligently to fill the long I/O queues
so that we hide the performance imbalance among SSDs even when there are
few parallel application
writes. We further define a policy of selecting dirty pages to flush and
a policy of taking out stale flush requests to reduce the amount of data
written to SSDs. We evaluate our solution in a real system.
Experiments show that our solution fully utilizes all SSDs
in an array under random write-heavy workloads. It improves I/O
throughput by up to 62\% under random workloads of mixed reads and writes
when SSDs are under active garbage collection.
It causes little extra data writeback and increases the cache hit rate.

\section{Introduction}
Solid state disks (SSDs) achieve great success due to significant performance
improvement over traditional hard drives in random I/O. However, due to
hardware limitation,
SSDs require an expensive erase operation before writing data to used blocks.
The granularity of the erase operation is usually multiple pages.
To counter the cost of erase, most SSDs use a log
structure to organize data and have the Flash Translation Layer (FTL)
to map data to physical locations on an SSD. Thus, SSDs require garbage
collection to clean space after substantial data write. Heavy random writes
trigger frequent garbage collection and slow down SSDs.

Much effort has been made to reduce overhead of garbage collection in SSDs
\cite{BPLRU, SFS, IPL, DFTL, LAST} and SSD vendors also add much intelligence
to their firmware. They all achieve a certain degree of success, but
the overhead of garbage collection can never be eliminated completely.

In an SSD array, unsynchronized garbage collection in individual SSDs leads
to performance degradation. Due to the unsynchronized garbage collection,
SSDs of the same model have
different throughput at any particular moment. Both hardware RAID controllers
and the software RAID in the Linux kernel only allow a limited number of pending
I/O requests. As a result, even though the I/O queue in the RAID controller
or the software RAID is filled with requests, some SSDs may still starve for
requests. Such performance imbalance among SSDs leads to underutilization of
some of the SSDs.

A possible solution is to synchronize garbage collection among SSDs.
Such a solution requires extra hardware added to SSDs and RAID controllers,
as suggested by Kim et al. \cite{Harmonia}.
Therefore, it requires coordination of SSD vendors and RAID
controller vendors. It can hardly become reality and benefit end users
in a short future.

We propose a software solution to tackle the unsynchronized garbage
collection in an SSD array and implement our solution in the set-associative
filesystem (SAFS) \cite{SAFS}, designed to provide maximal performance of
an SSD array. It is a general solution and does not rely on
any specific SSD characteristics. Instead of using RAID controllers, we attach
SSDs to host bus adapters (HBA) and expose individual SSDs to an operating
system. 
We maintain a short high-priority I/O queue for application requests and
a long low-priority I/O queue for flush requests in the main memory for each SSD.
The short high-priority I/O queues keep the latency of application I/O
requests low, while the long low-priority I/O queues hide the performance
imbalance among SSDs caused by garbage collection. We utilize the page
cache in SAFS to absorb application writes and design a flushing scheme
to write back dirty pages intelligently. We
further define a policy of selecting dirty pages to flush and
a policy of taking out stale flush requests to reduce the amount of data
written to SSDs.

Experiments show promising results. The design fully utilizes all SSDs in
an array and improves the performance of SAFS under random write-heavy
workloads. It increases the I/O throughput of SAFS by up to 64\% under mixed
read/write workloads. The design increases
the cache hit rate and flushes insignificant amount of extra data to SSDs.


\section{Related Work}
There is enormous amount of work on reducing overhead of garbage collection
on a single SSD. For instance, SFS \cite{SFS} is a file system specifically
designed for SSDs
to reduce overhead of garbage collection. It groups data blocks with similar
update likelihood into the same segments to reduce the amount of data copied in
garbage collection. BPLRU \cite{BPLRU} is a buffer management scheme for
the firmware inside SSDs. It uses a block-level LRU to manage the write buffer,
and a page padding technique when flushing victim blocks. In-page logging (IPL)
\cite{IPL} is a buffer management scheme designed for DBMS. It reserves some
space in each erase block of an SSD to log small writes to the block and
reconstructs data for reads.
Our solution works on multiple SSDs and treats each SSD as a black box, so it
can be well integrated with these techniques.

Kim et al. \cite{Harmonia} suggested to build an SSD-aware RAID controller and
SSD devices capable of global garbage collection to synchronize garbage
collection in an SSD array. Their solution requires the advance of both
SSD devices and RAID controllers and they evaluation their solution with simulation.
In contrast, we provide a software solution for commodity hardware and have
an implementation in a real system for evaluation. It benefits
users immediately.

\section{Design}
Our solution extends our previous work on SAFS \cite{SAFS}, a user-space
filesystem designed to achieve maximal performance from an SSD array
in a NUMA machine, to tackle unsynchronized garbage collection in an SSD array.
The root of inefficiency in an SSD RAID under garbage collection is
the limited size of the I/O queue of the RAID. The SSDs under active garbage
collection cannot keep up with other SSDs in the RAID and the overall performance
of the SSD RAID is limited by the slowest SSD. Therefore, our solution increases I/O
queues in SAFS and deploys a dirty page flusher to achieve maximal performance
from an SSD array
with a small number of parallel I/Os. Currently, we implement our solution
in the user space.


\subsection{Architecture}

\begin{figure}[t]
\centering
\includegraphics[scale=0.7]{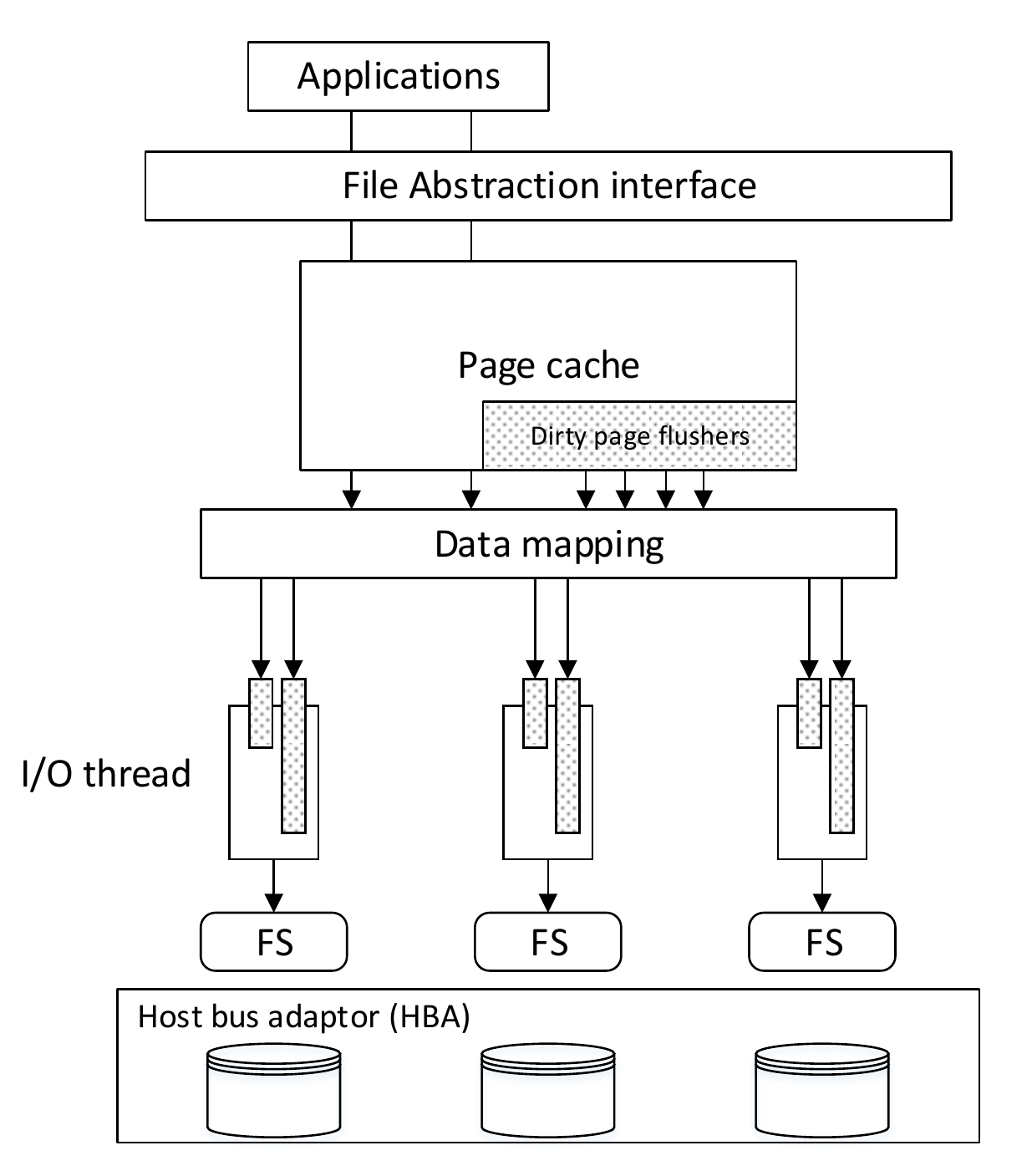}
\vspace{-5pt}
\caption{The architecture of SAFS. The shaded components of SAFS are modified
to tackle unsynchronized garbage collection.}
\vspace{-5pt}
\label{arch}
\end{figure}

The architecture of SAFS in Figure \ref{arch} has five components:
the file abstraction interface, the page cache, the data mapping layer,
I/O queues and I/O threads. SAFS exposes a file abstraction
interface to applications to receive I/O requests and notify the applications
of the completion of requests. Currently, it supports an asynchronous I/O
interface. SAFS is equipped with a light-weight, scalable page cache
called SA-cache \cite{hotstorage12}, where pages are grouped into many small page sets.
As shown by Zheng et al. \cite{hotstorage12}, Linux page cache has very high locking
overhead in a large parallel machine when the page turnover rate is high, due to
the global locks on the page cache. By grouping pages into many small page sets,
SA-cache eliminates the locking overhead.
Beneath the page cache is the data mapping
layer, which splits and dispatches I/O requests to SSDs. SSDs are connected to
the machine via host bus adapters (HBA), thus individual SSDs are exposed to
the operating system. Each SSD has a native filesystem to manage the data stored
on the SSD. It also has a dedicated I/O thread and originally has only one
dedicated I/O queue to buffer I/O requests.
Concurrent access to SSDs causes significant lock contention in the block subsystem
of an operating system. The dedicated I/O threads reduce the lock contention
in the operating system when issuing I/O requests to SSDs.

To tackle unsynchronized garbage collection in an SSD array, we modify
the I/O queues associated with SSDs and add a dirty page flusher to
the page cache of SAFS,
shown as the shaded components in Figure \ref{arch}. We split the original
I/O queue of an SSD into
two queues: a short high-priority queue and a long low-priority queue.
The dirty page flusher pre-cleans dirty pages in the page cache and issues
parallel write requests to SSDs.

\subsection{I/O queues and prioritized I/O requests}

SAFS \cite{SAFS} maintains an I/O queue for each SSD in the main memory,
and these I/O queues can be made substantially large to hide performance disparity
among SSDs. When some SSDs stall due to active garbage collection, application
requests can still be dispatched to any I/O queue. Therefore, applications
are not blocked by the garbage collection in some SSDs.

However, simply increasing the length of I/O queues cannot completely solve
the problem. Only applications capable of issuing many parallel I/O requests
can benefit from the large I/O queues. Therefore, we flush dirty pages
in the page cache to fill the long I/O queues. In a mixed read/write workload,
the I/O queues are filled with application read requests and flush requests.
It leads to long latency in application reads.

The solution is to split each I/O queue into two queues to provide
different service quality for different types of I/O requests.
One contains high-priority interactive I/O requests (application reads)
and the other contains low-priority background I/O requests (flush requests).
Only when there are no high-priority requests, the I/O threads issue
the low-priority requests to SSDs. Hence, application reads get much
shorter service time. It is essential to reduce the service time for
application reads in the case of read-update-write. Any unaligned write
requires read-update-write. Reducing the service time of reads allows
applications to perform read-update-writes at a higher rate.

To further reduce the service time of application requests, we always reserve
some I/O slots on
each SSD for application requests even if there are no application requests
at a moment. Such a decision is made based on the fact that SSDs can run at
decent performance even if they do not receive the maximal number of parallel
I/O requests required by the SSDs. When application requests are added to
the high-priority queue, they
are issued to SSDs immediately. An SSD typically requires 32 parallel I/O
requests to achieve maximal performance and we empirically reserve seven I/O
slots for high-priority I/O requests.



\subsection{A dirty page flusher}

The task of the dirty page flusher is to issue many flush requests to fill
the I/O queues while keeping the amount of data written back to SSDs small.
Filling the I/O queues with flush requests potentially leads to writing
much more data to SSDs than necessary. It is essentially important to
reduce data writeback because it helps increase
the application-perceived I/O throughput and reduce SSD wear-out.

The set-associative cache in SAFS composes of many small page sets and
the dirty page flusher is triggered to write back dirty pages in page sets
where the number of dirty pages exceeds a threshold. We empirically set the size
of a page set to 12 \cite{SAFS} and set the threshold to 6.
The flusher writes back only a small number of (one or two) dirty pages from
a page set each time. A page set that contains more dirty pages for writing back
will be placed in a FIFO queue. Once some flush requests complete, the flusher
checks the page sets in the queue in a round-robin manner and issue more flush
requests until no pages can be flushed in the page sets.
The algorithm gives each page set a chance to flush dirty pages but
is biased in favor of the page sets that get more writes.

The dirty page flusher together with the page cache reduces the average
latency of application writes, thus dramatically reducing
the number of parallel application writes required to achieve good performance.
When application writes hit page cache, they return immediately if
the required pages exist in the cache. In the case of cache misses, writes
can also return immediately if the evicted pages are clean. Application
writes may trigger page writes to SSDs if the victim page is dirty,
and they have to wait until the page writes to SSDs complete.
With the help of the dirty page flusher, the page cache maintains a certain
number of clean pages. Therefore, majority of writes are absorbed by
the page cache and return immediately. To further reduce the latency of
application writes, we tweak page eviction policies in SAFS to favor evicting
clean pages, similar to clean-first LRU \cite{CFLRU}.

Clean-first page eviction policies may reduce the cache hit rate, as they
ignore dirty pages when clean pages exist, and the dirty page flusher
alleviates the problem. The dirty page flusher writes back dirty pages that
are most likely to be evicted based on the page eviction policy.
Once the data of a dirty page is written back to an SSD, the page is likely
to be evicted. As a result, we essentially run the page eviction policy
on clean pages and dirty pages separately. More sophisticated cache management
policies such as \cite{FOR, LRU-WSR} may be used to better balance read and write
performance.

The flush requests in a long low-priority I/O queue are subject to long
latency and may be discarded. Given the length of a low-priority queue,
a flush request may take a long time to reach the head of the queue.
When it does, it may have become stale because the page in the request may
have been written back
to SSDs or is no longer urgent to be flushed based on the page eviction policy.
It is computationally expensive to sort all flush requests
in the I/O queue to find the most urgent
ones to flush. Instead, we simply discard all stale flush requests,
which gives more urgent flush requests a better chance to be written to SSDs.
Once discarding stale flush requests, an I/O thread will notify
the page cache and ask for more flush requests.
The scheme of discarding stale flush requests ensures that most flush requests
written to SSDs are needed to be flushed regardless of the length of
the I/O queues.

The minimal number of parallel flush requests required to hide the speed
disparity in an SSD array depends on the hardware configuration of the SSD
array. Instead of measuring and setting the minimal number for each SSD
array configuration, we only require users to loosely set a maximal number
of pending flush requests for an SSD array to avoid having too many flush
requests in the queue.
We empirically set the maximal number of pending flush requests to
$2048 \times$ the number of SSDs.


\subsubsection{Policy of selecting dirty pages for flushing} \label{predict}

The dirty page flusher executes a policy of selecting dirty pages inside
each page set. The policy iterates all pages in a page set and assigns
a flush score to each page. Thanks to the small size of a page set, there
is only small overhead in iterating all pages and computing scores.
The current implementation computes scores based on a page eviction policy,
given the fact that a dirty page that is more likely to be evicted is more
urgent to be flushed to SSDs. The pages that are more likely to be evicted
get higher flush scores.

We compute the flush score for GClock \cite{GCLOCK}, one of the page eviction
policies supported by SAFS, as follows. We first compute a distance score
for each page based on the number of hits and the distance to the clock head.

$distance\_score = hits \times set\_size + distance$

\noindent We sort the pages based on the distance scores and use the rank of a page
in the sorted array as a flush score. The pages with lower distance scores
get higher flush scores.

\subsubsection{Policy of discarding flush requests}

An I/O thread discards flush requests with the following policies: (i) the page
in the flush request has been evicted; (ii) the page in the flush request has
been cleaned; (iii) the page in the flush request has a flush score lower than
a threshold. Discarding flush requests with low flush scores
avoids the pages that are likely to be accessed in the future from being
evicted by the clean-first page eviction policy.

\subsection{Discussion}

The flushing scheme maximizes the write throughput but potentially reorders
write requests. Therefore, it benefits the applications that allow write
reordering. For applications that have more restrict write ordering, we
need to introduce a write barrier to SAFS to ensure all writes before the barrier
have been written to SSDs. Issuing write barriers frequently diminishes
the benefit of the flushing scheme. The applications that require very
strict write ordering can hardly benefit from the flushing scheme.

\section{Evaluation}
We evaluate our design on a non-uniform memory architecture machine with four
Intel Xeon E5-4620 processors, clocked at 2.2GHz, and 512GB memory of
DDR3-1333. Each processor has eight cores and have hyperthreads enabled.
The machine has three LSI HBA controllers connected to a SuperMicro storage
chassis, where 18 OCZ Vertex 4 SSDs are installed. Each SSD has 128GB.
The machine runs Ubuntu Server 12.04 and Linux kernel v3.2.30.

Due to the complex internal structure and firmware, an SSD may show
different performance in different runs even under the same workload.
To stablize the I/O performance
of an SSD, we write a large amount of data sequentially to the SSD
and keep it idle for 10 minutes before each experiment.  
All I/O throughput is measured when garbage collection on SSDs becomes
active.


\subsection{Impact of garbage collection}

We first explore the impact of garbage collection on an SSD and an SSD array.
We conduct experiments with random workloads to explore the following questions:
\begin{itemize}
	\item Question 1: how does disk occupancy of an SSD affect garbage collection?
	\item Question 2: how does the number of SSDs in an array affect the throughput
		of the array when garbage collection becomes active?
	\item Question 3: what is the minimal number of parallel writes required to achieve
		the maximal throughput of an SSD array under active garbage collection?
\end{itemize}

For question 1, we conduct experiments that write 60GB with 4KB uniformly
random writes to an SSD and show the result in Table \ref{occupancy}.
We measure I/O throughput when the SSD is 40\%, 60\%, 80\% full.
Table \ref{occupancy} shows that when garbage collection becomes active, the SSD
filled with more data has lower I/O throughput. It means that garbage
collection becomes more active when an SSD is filled with more data.
Garbage collection affects write throughput in all tests.

For question 2, we conduct experiments that dump data to 6, 12, 18 SSDs,
attached to 1, 2, 3 HBAs, respectively, and the result is shown in Table
\ref{scale}. Each experiment writes 40GB to each SSD with 4KB random writes.
All SSDs are 60\% full, and each SSD allows to have 128 pending I/O requests. 
Table \ref{scale} shows that the I/O throughput of each
individual SSD decreases as the number of SSDs in the array increases.
The result is expected. When more SSDs are installed in the array, more
SSDs can interfere the performance of the array. We expect the performance
of the array will further decrease when more SSDs are installed.

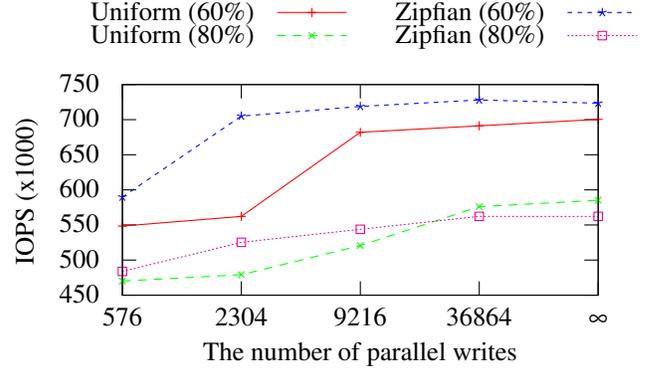
\begin{figure}[tb]
	\begin{center}
		\vspace{-15pt}
		\begin{tikzpicture}[gnuplot]
\path (0.000,0.000) rectangle (8.382,5.080);
\gpcolor{color=gp lt color border}
\gpsetlinetype{gp lt border}
\gpsetlinewidth{1.00}
\draw[gp path] (1.504,0.985)--(1.684,0.985);
\draw[gp path] (7.829,0.985)--(7.649,0.985);
\node[gp node right] at (1.320,0.985) { 450};
\draw[gp path] (1.504,1.452)--(1.684,1.452);
\draw[gp path] (7.829,1.452)--(7.649,1.452);
\node[gp node right] at (1.320,1.452) { 500};
\draw[gp path] (1.504,1.919)--(1.684,1.919);
\draw[gp path] (7.829,1.919)--(7.649,1.919);
\node[gp node right] at (1.320,1.919) { 550};
\draw[gp path] (1.504,2.386)--(1.684,2.386);
\draw[gp path] (7.829,2.386)--(7.649,2.386);
\node[gp node right] at (1.320,2.386) { 600};
\draw[gp path] (1.504,2.853)--(1.684,2.853);
\draw[gp path] (7.829,2.853)--(7.649,2.853);
\node[gp node right] at (1.320,2.853) { 650};
\draw[gp path] (1.504,3.320)--(1.684,3.320);
\draw[gp path] (7.829,3.320)--(7.649,3.320);
\node[gp node right] at (1.320,3.320) { 700};
\draw[gp path] (1.504,3.787)--(1.684,3.787);
\draw[gp path] (7.829,3.787)--(7.649,3.787);
\node[gp node right] at (1.320,3.787) { 750};
\draw[gp path] (1.504,0.985)--(1.504,1.165);
\draw[gp path] (1.504,3.787)--(1.504,3.607);
\node[gp node center] at (1.504,0.677) {576};
\draw[gp path] (3.085,0.985)--(3.085,1.165);
\draw[gp path] (3.085,3.787)--(3.085,3.607);
\node[gp node center] at (3.085,0.677) {2304};
\draw[gp path] (4.667,0.985)--(4.667,1.165);
\draw[gp path] (4.667,3.787)--(4.667,3.607);
\node[gp node center] at (4.667,0.677) {9216};
\draw[gp path] (6.248,0.985)--(6.248,1.165);
\draw[gp path] (6.248,3.787)--(6.248,3.607);
\node[gp node center] at (6.248,0.677) {36864};
\draw[gp path] (7.829,0.985)--(7.829,1.165);
\draw[gp path] (7.829,3.787)--(7.829,3.607);
\node[gp node center] at (7.829,0.677) {$\infty$};
\draw[gp path] (1.504,3.787)--(1.504,0.985)--(7.829,0.985)--(7.829,3.787)--cycle;
\node[gp node center,rotate=-270] at (0.246,2.386) {IOPS (x1000)};
\node[gp node center] at (4.666,0.215) {The number of parallel writes};
\node[gp node right] at (3.382,4.746) {Uniform (60\%)};
\gpcolor{color=gp lt color 0}
\gpsetlinetype{gp lt plot 0}
\draw[gp path] (3.566,4.746)--(4.482,4.746);
\draw[gp path] (1.504,1.904)--(3.085,2.033)--(4.667,3.152)--(6.248,3.238)--(7.829,3.324);
\gpsetpointsize{4.00}
\gppoint{gp mark 1}{(1.504,1.904)}
\gppoint{gp mark 1}{(3.085,2.033)}
\gppoint{gp mark 1}{(4.667,3.152)}
\gppoint{gp mark 1}{(6.248,3.238)}
\gppoint{gp mark 1}{(7.829,3.324)}
\gppoint{gp mark 1}{(4.024,4.746)}
\gpcolor{color=gp lt color border}
\node[gp node right] at (3.382,4.438) {Uniform (80\%)};
\gpcolor{color=gp lt color 1}
\gpsetlinetype{gp lt plot 1}
\draw[gp path] (3.566,4.438)--(4.482,4.438);
\draw[gp path] (1.504,1.172)--(3.085,1.258)--(4.667,1.645)--(6.248,2.162)--(7.829,2.248);
\gppoint{gp mark 2}{(1.504,1.172)}
\gppoint{gp mark 2}{(3.085,1.258)}
\gppoint{gp mark 2}{(4.667,1.645)}
\gppoint{gp mark 2}{(6.248,2.162)}
\gppoint{gp mark 2}{(7.829,2.248)}
\gppoint{gp mark 2}{(4.024,4.438)}
\gpcolor{color=gp lt color border}
\node[gp node right] at (7.242,4.746) {Zipfian (60\%)};
\gpcolor{color=gp lt color 2}
\gpsetlinetype{gp lt plot 2}
\draw[gp path] (7.426,4.746)--(8.342,4.746);
\draw[gp path] (1.504,2.291)--(3.085,3.367)--(4.667,3.496)--(6.248,3.582)--(7.829,3.539);
\gppoint{gp mark 3}{(1.504,2.291)}
\gppoint{gp mark 3}{(3.085,3.367)}
\gppoint{gp mark 3}{(4.667,3.496)}
\gppoint{gp mark 3}{(6.248,3.582)}
\gppoint{gp mark 3}{(7.829,3.539)}
\gppoint{gp mark 3}{(7.884,4.746)}
\gpcolor{color=gp lt color border}
\node[gp node right] at (7.242,4.438) {Zipfian (80\%)};
\gpcolor{color=gp lt color 3}
\gpsetlinetype{gp lt plot 3}
\draw[gp path] (7.426,4.438)--(8.342,4.438);
\draw[gp path] (1.504,1.301)--(3.085,1.688)--(4.667,1.861)--(6.248,2.033)--(7.829,2.033);
\gppoint{gp mark 4}{(1.504,1.301)}
\gppoint{gp mark 4}{(3.085,1.688)}
\gppoint{gp mark 4}{(4.667,1.861)}
\gppoint{gp mark 4}{(6.248,2.033)}
\gppoint{gp mark 4}{(7.829,2.033)}
\gppoint{gp mark 4}{(7.884,4.438)}
\gpcolor{color=gp lt color border}
\gpsetlinetype{gp lt border}
\draw[gp path] (1.504,3.787)--(1.504,0.985)--(7.829,0.985)--(7.829,3.787)--cycle;
\gpdefrectangularnode{gp plot 1}{\pgfpoint{1.504cm}{0.985cm}}{\pgfpoint{7.829cm}{3.787cm}}
\end{tikzpicture}
		\vspace{-15pt}
		\caption{The I/O throughput of 4KB random write to an array of 18 SSDs with
			different numbers of parallel writes under uniformly random and Zipfian
			random workloads.}
		\label{queue_lens}
	\end{center}
\end{figure}

For question 3, we conduct experiments that write data to 18 SSDs under
uniformly random and the Zipfian write-only workloads and vary the number
of parallel writes.
Figure \ref{queue_lens} shows that the I/O throughput increases by up to 28\%
when the number of parallel writes increases. With a sufficiently large
number of parallel writes, we can eventually reach the same
performance as each SSD being accessed independently. I/O access patterns can affect
the number of parallel writes required to achieve good throughput. Zipfian
random workloads require 2304 parallel writes in the SSD array to reach
approximately 95\% of maximal throughput.
In contrast, uniformly random workloads need 9216 parallel writes or even more.
Nevertheless, we need to use thousands of or tens of thousands of parallel
writes to hide the speed disparity of individual SSDs caused by garbage
collection. Based on this experiment and the previous one, we
expect that the number of parallel writes required to achieve
good performance increases super-linearly with the number of SSDs in an array.

\begin{table}
	\begin{center}
		\small
		\begin{tabular}{|c|c|c|c|c|}
			\hline
			Occupancy & maximal & 40\% & 60\% & 80\% \\
			\hline
			IOPS & 60928 & 42240 & 38656 & 32512 \\
			\hline
		\end{tabular}
		\normalsize
	\end{center}
	\caption{The I/O throughput of 4KB random write to an SSD with different disk
	occupancy under active garbage collection. The maximal throughput is measured
	when there is no garbage collection.}
	\label{occupancy}
\end{table}



\begin{table}
	\begin{center}
		\small
		\begin{tabular}{|c|c|c|c|c|}
			\hline
			The number of SSDs & 1 & 6 & 12 & 18 \\
			\hline
			IOPS per SSD & 38656 & 37888 & 33280 & 31744 \\
			\hline
		\end{tabular}
		\normalsize
	\end{center}
	\caption{The average I/O throughput of 4KB random write per SSD in arrays
		of different sizes when each SSD is under active garbage collection.
	The number of parallel writes per SSD is 128.}
	\label{scale}
\end{table}

\subsection{Effectiveness of the dirty page flusher}

We measure the effectiveness of the dirty page flusher by benchmarking SAFS
under uniformly random write workloads and Zipfian random
write workloads with and without the dirty page flusher enabled.
We measure the I/O throughput improved by the dirty page flusher, as well as
the amount of extra data writeback caused by the flusher and the cache hit rate.
We evaluate both synchronous and asynchronous I/O. Asynchronous I/O
uses I/O depth of 32 per SSD. All SSDs are 80\% full.


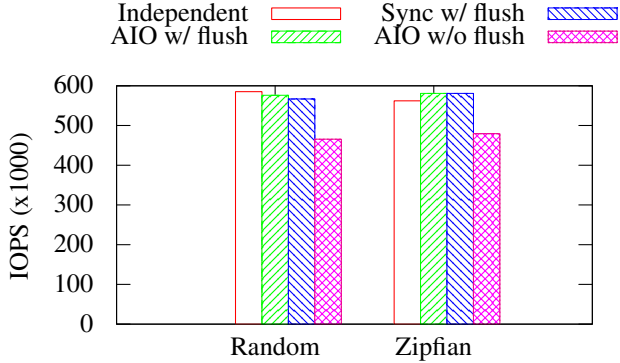
\begin{figure}[tb]
	\begin{center}
		\vspace{-15pt}
		\begin{tikzpicture}[gnuplot]
\path (0.000,0.000) rectangle (8.382,5.080);
\gpcolor{color=gp lt color border}
\gpsetlinetype{gp lt border}
\gpsetlinewidth{1.00}
\draw[gp path] (1.504,0.616)--(1.684,0.616);
\draw[gp path] (7.829,0.616)--(7.649,0.616);
\node[gp node right] at (1.320,0.616) { 0};
\draw[gp path] (1.504,1.145)--(1.684,1.145);
\draw[gp path] (7.829,1.145)--(7.649,1.145);
\node[gp node right] at (1.320,1.145) { 100};
\draw[gp path] (1.504,1.673)--(1.684,1.673);
\draw[gp path] (7.829,1.673)--(7.649,1.673);
\node[gp node right] at (1.320,1.673) { 200};
\draw[gp path] (1.504,2.202)--(1.684,2.202);
\draw[gp path] (7.829,2.202)--(7.649,2.202);
\node[gp node right] at (1.320,2.202) { 300};
\draw[gp path] (1.504,2.730)--(1.684,2.730);
\draw[gp path] (7.829,2.730)--(7.649,2.730);
\node[gp node right] at (1.320,2.730) { 400};
\draw[gp path] (1.504,3.259)--(1.684,3.259);
\draw[gp path] (7.829,3.259)--(7.649,3.259);
\node[gp node right] at (1.320,3.259) { 500};
\draw[gp path] (1.504,3.787)--(1.684,3.787);
\draw[gp path] (7.829,3.787)--(7.649,3.787);
\node[gp node right] at (1.320,3.787) { 600};
\draw[gp path] (3.612,0.616)--(3.612,0.796);
\draw[gp path] (3.612,3.787)--(3.612,3.607);
\node[gp node center] at (3.612,0.308) {Random};
\draw[gp path] (5.721,0.616)--(5.721,0.796);
\draw[gp path] (5.721,3.787)--(5.721,3.607);
\node[gp node center] at (5.721,0.308) {Zipfian};
\draw[gp path] (1.504,3.787)--(1.504,0.616)--(7.829,0.616)--(7.829,3.787)--cycle;
\node[gp node center,rotate=-270] at (0.246,2.201) {IOPS (x1000)};
\node[gp node right] at (3.382,4.746) {Independent};
\def\gpfillpath{(3.566,4.669)--(4.482,4.669)--(4.482,4.823)--(3.566,4.823)--cycle}
\gpfill{color=gpbgfillcolor} \gpfillpath;
\gpfill{color=gp lt color 0,gp pattern 0,pattern color=.} \gpfillpath;
\gpcolor{color=gp lt color 0}
\gpsetlinetype{gp lt plot 0}
\draw[gp path] (3.566,4.669)--(4.482,4.669)--(4.482,4.823)--(3.566,4.823)--cycle;
\def\gpfillpath{(3.085,0.616)--(3.438,0.616)--(3.438,3.710)--(3.085,3.710)--cycle}
\gpfill{color=gpbgfillcolor} \gpfillpath;
\gpfill{color=gp lt color 0,gp pattern 0,pattern color=.} \gpfillpath;
\draw[gp path] (3.085,0.616)--(3.085,3.709)--(3.437,3.709)--(3.437,0.616)--cycle;
\def\gpfillpath{(5.194,0.616)--(5.546,0.616)--(5.546,3.588)--(5.194,3.588)--cycle}
\gpfill{color=gpbgfillcolor} \gpfillpath;
\gpfill{color=gp lt color 0,gp pattern 0,pattern color=.} \gpfillpath;
\draw[gp path] (5.194,0.616)--(5.194,3.587)--(5.545,3.587)--(5.545,0.616)--cycle;
\gpcolor{color=gp lt color border}
\node[gp node right] at (3.382,4.438) {AIO w/ flush};
\def\gpfillpath{(3.566,4.361)--(4.482,4.361)--(4.482,4.515)--(3.566,4.515)--cycle}
\gpfill{color=gpbgfillcolor} \gpfillpath;
\gpfill{color=gp lt color 1,gp pattern 1,pattern color=.} \gpfillpath;
\gpcolor{color=gp lt color 1}
\gpsetlinetype{gp lt plot 1}
\draw[gp path] (3.566,4.361)--(4.482,4.361)--(4.482,4.515)--(3.566,4.515)--cycle;
\def\gpfillpath{(3.437,0.616)--(3.789,0.616)--(3.789,3.661)--(3.437,3.661)--cycle}
\gpfill{color=gpbgfillcolor} \gpfillpath;
\gpfill{color=gp lt color 1,gp pattern 1,pattern color=.} \gpfillpath;
\draw[gp path] (3.437,0.616)--(3.437,3.660)--(3.788,3.660)--(3.788,0.616)--cycle;
\def\gpfillpath{(5.545,0.616)--(5.897,0.616)--(5.897,3.686)--(5.545,3.686)--cycle}
\gpfill{color=gpbgfillcolor} \gpfillpath;
\gpfill{color=gp lt color 1,gp pattern 1,pattern color=.} \gpfillpath;
\draw[gp path] (5.545,0.616)--(5.545,3.685)--(5.896,3.685)--(5.896,0.616)--cycle;
\gpcolor{color=gp lt color border}
\node[gp node right] at (7.058,4.746) {Sync w/ flush};
\def\gpfillpath{(7.242,4.669)--(8.158,4.669)--(8.158,4.823)--(7.242,4.823)--cycle}
\gpfill{color=gpbgfillcolor} \gpfillpath;
\gpfill{color=gp lt color 2,gp pattern 2,pattern color=.} \gpfillpath;
\gpcolor{color=gp lt color 2}
\gpsetlinetype{gp lt plot 2}
\draw[gp path] (7.242,4.669)--(8.158,4.669)--(8.158,4.823)--(7.242,4.823)--cycle;
\def\gpfillpath{(3.788,0.616)--(4.140,0.616)--(4.140,3.612)--(3.788,3.612)--cycle}
\gpfill{color=gpbgfillcolor} \gpfillpath;
\gpfill{color=gp lt color 2,gp pattern 2,pattern color=.} \gpfillpath;
\draw[gp path] (3.788,0.616)--(3.788,3.611)--(4.139,3.611)--(4.139,0.616)--cycle;
\def\gpfillpath{(5.896,0.616)--(6.249,0.616)--(6.249,3.686)--(5.896,3.686)--cycle}
\gpfill{color=gpbgfillcolor} \gpfillpath;
\gpfill{color=gp lt color 2,gp pattern 2,pattern color=.} \gpfillpath;
\draw[gp path] (5.896,0.616)--(5.896,3.685)--(6.248,3.685)--(6.248,0.616)--cycle;
\gpcolor{color=gp lt color border}
\node[gp node right] at (7.058,4.438) {AIO w/o flush};
\def\gpfillpath{(7.242,4.361)--(8.158,4.361)--(8.158,4.515)--(7.242,4.515)--cycle}
\gpfill{color=gpbgfillcolor} \gpfillpath;
\gpfill{color=gp lt color 3,gp pattern 3,pattern color=.} \gpfillpath;
\gpcolor{color=gp lt color 3}
\gpsetlinetype{gp lt plot 3}
\draw[gp path] (7.242,4.361)--(8.158,4.361)--(8.158,4.515)--(7.242,4.515)--cycle;
\def\gpfillpath{(4.139,0.616)--(4.492,0.616)--(4.492,3.077)--(4.139,3.077)--cycle}
\gpfill{color=gpbgfillcolor} \gpfillpath;
\gpfill{color=gp lt color 3,gp pattern 3,pattern color=.} \gpfillpath;
\draw[gp path] (4.139,0.616)--(4.139,3.076)--(4.491,3.076)--(4.491,0.616)--cycle;
\def\gpfillpath{(6.248,0.616)--(6.600,0.616)--(6.600,3.150)--(6.248,3.150)--cycle}
\gpfill{color=gpbgfillcolor} \gpfillpath;
\gpfill{color=gp lt color 3,gp pattern 3,pattern color=.} \gpfillpath;
\draw[gp path] (6.248,0.616)--(6.248,3.149)--(6.599,3.149)--(6.599,0.616)--cycle;
\gpcolor{color=gp lt color border}
\gpsetlinetype{gp lt border}
\draw[gp path] (1.504,3.787)--(1.504,0.616)--(7.829,0.616)--(7.829,3.787)--cycle;
\gpdefrectangularnode{gp plot 1}{\pgfpoint{1.504cm}{0.616cm}}{\pgfpoint{7.829cm}{3.787cm}}
\end{tikzpicture}
		\vspace{-15pt}
		\caption{The I/O throughput of SAFS synchronous and asynchronous 4KB random
			write with and without the dirty page flusher under the uniformly random
			and Zipfian random workloads.
			We also include the throughput that all SSDs are written independently.}
		\label{compare}
	\end{center}
\end{figure}

\begin{figure}[tb]
	\begin{center}
		\vspace{-15pt}
		\begin{tikzpicture}[gnuplot]
\path (0.000,0.000) rectangle (8.382,4.572);
\gpcolor{color=gp lt color border}
\gpsetlinetype{gp lt border}
\gpsetlinewidth{1.00}
\draw[gp path] (1.504,0.616)--(1.684,0.616);
\draw[gp path] (7.829,0.616)--(7.649,0.616);
\node[gp node right] at (1.320,0.616) { 0};
\draw[gp path] (1.504,1.040)--(1.684,1.040);
\draw[gp path] (7.829,1.040)--(7.649,1.040);
\node[gp node right] at (1.320,1.040) { 50};
\draw[gp path] (1.504,1.465)--(1.684,1.465);
\draw[gp path] (7.829,1.465)--(7.649,1.465);
\node[gp node right] at (1.320,1.465) { 100};
\draw[gp path] (1.504,1.889)--(1.684,1.889);
\draw[gp path] (7.829,1.889)--(7.649,1.889);
\node[gp node right] at (1.320,1.889) { 150};
\draw[gp path] (1.504,2.314)--(1.684,2.314);
\draw[gp path] (7.829,2.314)--(7.649,2.314);
\node[gp node right] at (1.320,2.314) { 200};
\draw[gp path] (1.504,2.738)--(1.684,2.738);
\draw[gp path] (7.829,2.738)--(7.649,2.738);
\node[gp node right] at (1.320,2.738) { 250};
\draw[gp path] (1.504,3.163)--(1.684,3.163);
\draw[gp path] (7.829,3.163)--(7.649,3.163);
\node[gp node right] at (1.320,3.163) { 300};
\draw[gp path] (1.504,3.587)--(1.684,3.587);
\draw[gp path] (7.829,3.587)--(7.649,3.587);
\node[gp node right] at (1.320,3.587) { 350};
\draw[gp path] (3.612,0.616)--(3.612,0.796);
\draw[gp path] (3.612,3.587)--(3.612,3.407);
\node[gp node center] at (3.612,0.308) {Random};
\draw[gp path] (5.721,0.616)--(5.721,0.796);
\draw[gp path] (5.721,3.587)--(5.721,3.407);
\node[gp node center] at (5.721,0.308) {Zipfian};
\draw[gp path] (1.504,3.587)--(1.504,0.616)--(7.829,0.616)--(7.829,3.587)--cycle;
\node[gp node center,rotate=-270] at (0.246,2.101) {IOPS (x1000)};
\node[gp node right] at (3.382,4.238) {Without flush};
\def\gpfillpath{(3.566,4.161)--(4.482,4.161)--(4.482,4.315)--(3.566,4.315)--cycle}
\gpfill{color=gpbgfillcolor} \gpfillpath;
\gpfill{color=gp lt color 0,gp pattern 0,pattern color=.} \gpfillpath;
\gpcolor{color=gp lt color 0}
\gpsetlinetype{gp lt plot 0}
\draw[gp path] (3.566,4.161)--(4.482,4.161)--(4.482,4.315)--(3.566,4.315)--cycle;
\def\gpfillpath{(3.349,0.616)--(3.877,0.616)--(3.877,2.646)--(3.349,2.646)--cycle}
\gpfill{color=gpbgfillcolor} \gpfillpath;
\gpfill{color=gp lt color 0,gp pattern 0,pattern color=.} \gpfillpath;
\draw[gp path] (3.349,0.616)--(3.349,2.645)--(3.876,2.645)--(3.876,0.616)--cycle;
\def\gpfillpath{(5.457,0.616)--(5.985,0.616)--(5.985,2.671)--(5.457,2.671)--cycle}
\gpfill{color=gpbgfillcolor} \gpfillpath;
\gpfill{color=gp lt color 0,gp pattern 0,pattern color=.} \gpfillpath;
\draw[gp path] (5.457,0.616)--(5.457,2.670)--(5.984,2.670)--(5.984,0.616)--cycle;
\gpcolor{color=gp lt color border}
\node[gp node right] at (7.058,4.238) {With flush};
\def\gpfillpath{(7.242,4.161)--(8.158,4.161)--(8.158,4.315)--(7.242,4.315)--cycle}
\gpfill{color=gpbgfillcolor} \gpfillpath;
\gpfill{color=gp lt color 1,gp pattern 1,pattern color=.} \gpfillpath;
\gpcolor{color=gp lt color 1}
\gpsetlinetype{gp lt plot 1}
\draw[gp path] (7.242,4.161)--(8.158,4.161)--(8.158,4.315)--(7.242,4.315)--cycle;
\def\gpfillpath{(3.876,0.616)--(4.404,0.616)--(4.404,3.461)--(3.876,3.461)--cycle}
\gpfill{color=gpbgfillcolor} \gpfillpath;
\gpfill{color=gp lt color 1,gp pattern 1,pattern color=.} \gpfillpath;
\draw[gp path] (3.876,0.616)--(3.876,3.460)--(4.403,3.460)--(4.403,0.616)--cycle;
\def\gpfillpath{(5.984,0.616)--(6.512,0.616)--(6.512,3.427)--(5.984,3.427)--cycle}
\gpfill{color=gpbgfillcolor} \gpfillpath;
\gpfill{color=gp lt color 1,gp pattern 1,pattern color=.} \gpfillpath;
\draw[gp path] (5.984,0.616)--(5.984,3.426)--(6.511,3.426)--(6.511,0.616)--cycle;
\gpcolor{color=gp lt color border}
\gpsetlinetype{gp lt border}
\draw[gp path] (1.504,3.587)--(1.504,0.616)--(7.829,0.616)--(7.829,3.587)--cycle;
\gpdefrectangularnode{gp plot 1}{\pgfpoint{1.504cm}{0.616cm}}{\pgfpoint{7.829cm}{3.587cm}}
\end{tikzpicture}
		\vspace{-15pt}
		\caption{The I/O throughput of SAFS asynchronous write under
			uniformly random and Zipfian random workloads of unaligned writes.
			Each write is 128 bytes. We compare the throughput with and without
			the dirty page flusher.}
		\label{unaligned}
	\end{center}
\end{figure}
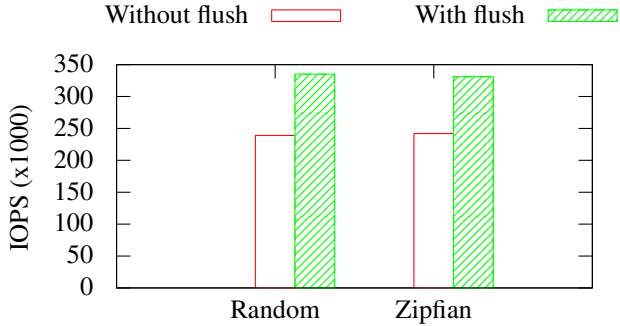

\begin{figure}[tb]
	\begin{center}
		\vspace{-15pt}
		\begin{tikzpicture}[gnuplot]
\path (0.000,0.000) rectangle (8.382,4.826);
\gpcolor{color=gp lt color border}
\gpsetlinetype{gp lt border}
\gpsetlinewidth{1.00}
\draw[gp path] (1.688,0.985)--(1.868,0.985);
\draw[gp path] (7.829,0.985)--(7.649,0.985);
\node[gp node right] at (1.504,0.985) { 0};
\draw[gp path] (1.688,1.504)--(1.868,1.504);
\draw[gp path] (7.829,1.504)--(7.649,1.504);
\node[gp node right] at (1.504,1.504) { 200};
\draw[gp path] (1.688,2.024)--(1.868,2.024);
\draw[gp path] (7.829,2.024)--(7.649,2.024);
\node[gp node right] at (1.504,2.024) { 400};
\draw[gp path] (1.688,2.543)--(1.868,2.543);
\draw[gp path] (7.829,2.543)--(7.649,2.543);
\node[gp node right] at (1.504,2.543) { 600};
\draw[gp path] (1.688,3.062)--(1.868,3.062);
\draw[gp path] (7.829,3.062)--(7.649,3.062);
\node[gp node right] at (1.504,3.062) { 800};
\draw[gp path] (1.688,3.581)--(1.868,3.581);
\draw[gp path] (7.829,3.581)--(7.649,3.581);
\node[gp node right] at (1.504,3.581) { 1000};
\draw[gp path] (2.565,0.985)--(2.565,1.165);
\draw[gp path] (2.565,3.841)--(2.565,3.661);
\node[gp node center] at (2.565,0.677) {100\%};
\draw[gp path] (3.443,0.985)--(3.443,1.165);
\draw[gp path] (3.443,3.841)--(3.443,3.661);
\node[gp node center] at (3.443,0.677) {80\%};
\draw[gp path] (4.320,0.985)--(4.320,1.165);
\draw[gp path] (4.320,3.841)--(4.320,3.661);
\node[gp node center] at (4.320,0.677) {60\%};
\draw[gp path] (5.197,0.985)--(5.197,1.165);
\draw[gp path] (5.197,3.841)--(5.197,3.661);
\node[gp node center] at (5.197,0.677) {40\%};
\draw[gp path] (6.074,0.985)--(6.074,1.165);
\draw[gp path] (6.074,3.841)--(6.074,3.661);
\node[gp node center] at (6.074,0.677) {20\%};
\draw[gp path] (6.952,0.985)--(6.952,1.165);
\draw[gp path] (6.952,3.841)--(6.952,3.661);
\node[gp node center] at (6.952,0.677) {0\%};
\draw[gp path] (1.688,3.841)--(1.688,0.985)--(7.829,0.985)--(7.829,3.841)--cycle;
\node[gp node center,rotate=-270] at (0.246,2.413) {IOPS (x1000)};
\node[gp node center] at (4.758,0.215) {Read percentage};
\node[gp node right] at (3.474,4.492) {Without flush};
\def\gpfillpath{(3.658,4.415)--(4.574,4.415)--(4.574,4.569)--(3.658,4.569)--cycle}
\gpfill{color=gpbgfillcolor} \gpfillpath;
\gpfill{color=gp lt color 0,gp pattern 0,pattern color=.} \gpfillpath;
\gpcolor{color=gp lt color 0}
\gpsetlinetype{gp lt plot 0}
\draw[gp path] (3.658,4.415)--(4.574,4.415)--(4.574,4.569)--(3.658,4.569)--cycle;
\def\gpfillpath{(2.456,0.985)--(2.676,0.985)--(2.676,3.551)--(2.456,3.551)--cycle}
\gpfill{color=gpbgfillcolor} \gpfillpath;
\gpfill{color=gp lt color 0,gp pattern 0,pattern color=.} \gpfillpath;
\draw[gp path] (2.456,0.985)--(2.456,3.550)--(2.675,3.550)--(2.675,0.985)--cycle;
\def\gpfillpath{(3.333,0.985)--(3.553,0.985)--(3.553,2.803)--(3.333,2.803)--cycle}
\gpfill{color=gpbgfillcolor} \gpfillpath;
\gpfill{color=gp lt color 0,gp pattern 0,pattern color=.} \gpfillpath;
\draw[gp path] (3.333,0.985)--(3.333,2.802)--(3.552,2.802)--(3.552,0.985)--cycle;
\def\gpfillpath{(4.210,0.985)--(4.431,0.985)--(4.431,2.585)--(4.210,2.585)--cycle}
\gpfill{color=gpbgfillcolor} \gpfillpath;
\gpfill{color=gp lt color 0,gp pattern 0,pattern color=.} \gpfillpath;
\draw[gp path] (4.210,0.985)--(4.210,2.584)--(4.430,2.584)--(4.430,0.985)--cycle;
\def\gpfillpath{(5.087,0.985)--(5.308,0.985)--(5.308,2.053)--(5.087,2.053)--cycle}
\gpfill{color=gpbgfillcolor} \gpfillpath;
\gpfill{color=gp lt color 0,gp pattern 0,pattern color=.} \gpfillpath;
\draw[gp path] (5.087,0.985)--(5.087,2.052)--(5.307,2.052)--(5.307,0.985)--cycle;
\def\gpfillpath{(5.965,0.985)--(6.185,0.985)--(6.185,1.965)--(5.965,1.965)--cycle}
\gpfill{color=gpbgfillcolor} \gpfillpath;
\gpfill{color=gp lt color 0,gp pattern 0,pattern color=.} \gpfillpath;
\draw[gp path] (5.965,0.985)--(5.965,1.964)--(6.184,1.964)--(6.184,0.985)--cycle;
\def\gpfillpath{(6.842,0.985)--(7.062,0.985)--(7.062,2.186)--(6.842,2.186)--cycle}
\gpfill{color=gpbgfillcolor} \gpfillpath;
\gpfill{color=gp lt color 0,gp pattern 0,pattern color=.} \gpfillpath;
\draw[gp path] (6.842,0.985)--(6.842,2.185)--(7.061,2.185)--(7.061,0.985)--cycle;
\gpcolor{color=gp lt color border}
\node[gp node right] at (7.150,4.492) {With flush};
\def\gpfillpath{(7.334,4.415)--(8.250,4.415)--(8.250,4.569)--(7.334,4.569)--cycle}
\gpfill{color=gpbgfillcolor} \gpfillpath;
\gpfill{color=gp lt color 1,gp pattern 1,pattern color=.} \gpfillpath;
\gpcolor{color=gp lt color 1}
\gpsetlinetype{gp lt plot 1}
\draw[gp path] (7.334,4.415)--(8.250,4.415)--(8.250,4.569)--(7.334,4.569)--cycle;
\def\gpfillpath{(2.675,0.985)--(2.895,0.985)--(2.895,3.551)--(2.675,3.551)--cycle}
\gpfill{color=gpbgfillcolor} \gpfillpath;
\gpfill{color=gp lt color 1,gp pattern 1,pattern color=.} \gpfillpath;
\draw[gp path] (2.675,0.985)--(2.675,3.550)--(2.894,3.550)--(2.894,0.985)--cycle;
\def\gpfillpath{(3.552,0.985)--(3.773,0.985)--(3.773,2.983)--(3.552,2.983)--cycle}
\gpfill{color=gpbgfillcolor} \gpfillpath;
\gpfill{color=gp lt color 1,gp pattern 1,pattern color=.} \gpfillpath;
\draw[gp path] (3.552,0.985)--(3.552,2.982)--(3.772,2.982)--(3.772,0.985)--cycle;
\def\gpfillpath{(4.430,0.985)--(4.650,0.985)--(4.650,2.793)--(4.430,2.793)--cycle}
\gpfill{color=gpbgfillcolor} \gpfillpath;
\gpfill{color=gp lt color 1,gp pattern 1,pattern color=.} \gpfillpath;
\draw[gp path] (4.430,0.985)--(4.430,2.792)--(4.649,2.792)--(4.649,0.985)--cycle;
\def\gpfillpath{(5.307,0.985)--(5.527,0.985)--(5.527,2.713)--(5.307,2.713)--cycle}
\gpfill{color=gpbgfillcolor} \gpfillpath;
\gpfill{color=gp lt color 1,gp pattern 1,pattern color=.} \gpfillpath;
\draw[gp path] (5.307,0.985)--(5.307,2.712)--(5.526,2.712)--(5.526,0.985)--cycle;
\def\gpfillpath{(6.184,0.985)--(6.404,0.985)--(6.404,2.331)--(6.184,2.331)--cycle}
\gpfill{color=gpbgfillcolor} \gpfillpath;
\gpfill{color=gp lt color 1,gp pattern 1,pattern color=.} \gpfillpath;
\draw[gp path] (6.184,0.985)--(6.184,2.330)--(6.403,2.330)--(6.403,0.985)--cycle;
\def\gpfillpath{(7.061,0.985)--(7.282,0.985)--(7.282,2.443)--(7.061,2.443)--cycle}
\gpfill{color=gpbgfillcolor} \gpfillpath;
\gpfill{color=gp lt color 1,gp pattern 1,pattern color=.} \gpfillpath;
\draw[gp path] (7.061,0.985)--(7.061,2.442)--(7.281,2.442)--(7.281,0.985)--cycle;
\gpcolor{color=gp lt color border}
\gpsetlinetype{gp lt border}
\draw[gp path] (1.688,3.841)--(1.688,0.985)--(7.829,0.985)--(7.829,3.841)--cycle;
\gpdefrectangularnode{gp plot 1}{\pgfpoint{1.688cm}{0.985cm}}{\pgfpoint{7.829cm}{3.841cm}}
\end{tikzpicture}
		\vspace{-15pt}
		\caption{The I/O throughput of SAFS asynchronous I/O under
			the uniformly random workloads with different read/write ratios.
			Each read/write is 4KB.}
		\label{mixed}
	\end{center}
\end{figure}
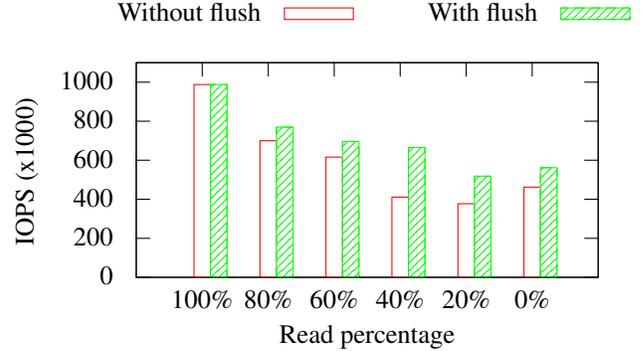

\begin{table}
	\begin{center}
		\small
		\begin{tabular}{|c|c|c|c|c|c|}
			\hline
			Read percentage & 80\% & 60\% & 40\% & 20\% & 0\% \\
			\hline
			Extra writeback & 2.4\% & 1.6\% & 2.2\% & 2.7\% & 3.2\% \\
			\hline
			Cache hit increase & 0.7\% & 0.6\% & 1\% & 1.4\% & 4\% \\
			\hline
		\end{tabular}
		\normalsize
	\end{center}
	\caption{The amount of extra dirty data writeback and the improvement of
	cache hit rate by the dirty page flusher under Zipfian random workloads
	with different read/write ratios, compared with cached I/O without
	the dirty page flusher. Each read/write is 4KB.}
	\label{zipfian}
\end{table}

We measure the I/O throughput of asynchronous writes and synchronous writes
under write-only random workloads. Figure \ref{compare} shows the I/O throughput
of aligned random writes.
When the dirty page flusher is enabled, both synchronous and asynchronous
writes can achieve maximal performance (when data is written to SSDs independently),
and improve the I/O throughput by up to 24\% than that without the dirty page flusher.
Figure \ref{unaligned} shows the I/O throughput of unaligned
random write. Each write triggers a page read from the SSD array, so synchronous
I/O cannot achieve good performance and is not shown in Figure \ref{unaligned}.
The dirty page flusher can improve
I/O throughput of asynchronous write by up to 39\%.

We measure the I/O throughput of asynchronous I/O under the uniformly random
workloads with different
read/write ratios (Figure \ref{mixed}). The dirty page flusher effectively
cleans up dirty pages and writes them back to faster SSDs when some SSDs are slowed
down by garbage collection. When garbage collection ceases, the page cache absorbs
writes and give reads more opportunity to be issued to SSDs. The flusher
improves I/O throughput even when read percentage is high. The largest
improvement occurs at read percentage of 40\%. The read/write throughput is
improved by 62\%.

We measure the amount of extra data written back and the cache hit rate
affected by the dirty page flusher under Zipfian random workloads with
different read/write ratios (Table \ref{zipfian}). We compare its result
with cached I/O without the dirty page flusher. Although the flusher
can cause extra data written back, the amount of extra writeback is fairly
small. Furthermore, the flushing scheme
slightly increases the cache hit rate because it helps evict dirty
pages that are unlikely to be accessed again.






\section{Conclusions}
We propose a software solution that tackles unsynchronized garbage collection
in an SSD array. We maintain long I/O queues in the main memory for each
SSD and use a dirty page flusher to pre-clean dirty pages and fill the long
I/O queues. We define a policy of selecting dirty pages to flush
and a policy of discarding stale flush requests to reduce the amount of data
flushed to SSDs.

We evaluate the design with uniformly random and Zipfian random workloads.
The design improves the I/O throughput by up to 28\% under write-only workloads,
and by up to 62\% under uniformly random mixed read/write workloads. We further
demonstrate that the design causes little extra data written back to SSDs and
slightly improves the cache hit rate under Zipfian random workloads.

{\footnotesize \bibliographystyle{acm}
\bibliography{fast14}}


\end{document}